\documentclass[lettersize,journal]{IEEEtran}
\usepackage{amsmath,amsfonts}
\usepackage{algorithmic}
\usepackage{algorithm}
\usepackage{array}
\usepackage[caption=false,font=normalsize,labelfont=sf,textfont=sf]{subfig}
\usepackage{textcomp}
\usepackage{url}
\usepackage{verbatim}
\usepackage{graphicx}
\usepackage{fancybox}
\usepackage{cite}
\usepackage{tabularx}
\usepackage{tikz}
\usepackage{circuitikz}
\usetikzlibrary{positioning}
\usepackage{amsthm}
\usepackage{cuted} 
\usepackage{placeins} 
\usepackage{dblfloatfix} 
\newtheorem{definition}{Definition}[section]
\usepackage{amssymb}
\usepackage{xcolor}
\def\BibTeX{{\rm B\kern-.05em{\sc i\kern-.025em b}\kern-.08em
    T\kern-.1667em\lower.7ex\hbox{E}\kern-.125emX}}
    
\begin{document}

\title{Trajectory-Based Nonlinear Indices for Real-Time Monitoring and Quantification of Short-Term Voltage Stability }

\author{Mohammad~Almomani,~\IEEEmembership{graduate student member,~IEEE,}
        Muhammad~Sarwar,~\IEEEmembership{graduate student member,~IEEE,} and ~ Venkataramana~Ajjarapu,~\IEEEmembership{Fellow,~IEEE}


\IEEEauthorblockA{\textit{Department of Electrical \& Computer Engineering, Iowa State University, Ames, IA, USA} \\
Emails: \{mmomani, msarwar, vajjarap\}@iastate.edu}
}

\maketitle

\begin{abstract}
Existing short-term voltage stability (STVS) methods typically address either voltage oscillations or delayed voltage recovery; however, the coexistence of both phenomena has not been adequately covered in the literature. Moreover, existing real-time STVS assessment methods often provide only binary stability classifications. This paper proposes novel indices that enable early detection and quantify the degree of stability. The proposed method decomposes post-fault voltage trajectories using Empirical Mode Decomposition (EMD) into residual and oscillatory components. It then employs Lyapunov Exponents (LEs) to characterize the dynamic behavior of each component and evaluates the stability degree using Kullback–Leibler (KL) divergence by comparing the LEs of each component with those of a predefined critical signal. The proposed indices assess oscillatory stability significantly faster than the traditional LE method applied directly to the original signal. Specifically, they detect stability within 0.6 seconds after a fault, compared to approximately 10 seconds for the conventional LE approach. In addition, the delayed-recovery index can identify generator trips caused by over-excitation limits within 3 seconds, well before the actual trip occurs at approximately 20 seconds, thereby providing operators and controllers sufficient time to take preventive actions. Furthermore, thresholds are derived to distinguish between stable and unstable cases, offering a graded measure of the stability margin. Simulation studies on the Nordic test system under varying load conditions demonstrate the effectiveness of the proposed indices.
\end{abstract}

\section{Introduction}
\doublebox{\parbox{0.8\linewidth}{
\centering
This paper has been submitted to IEEE Transactions on Power Systems.
}}

\small

The IEEE/CIGRE Joint Task Force classifies power system stability based on three key aspects: the nature of stability, the type of disturbance, and the time frame of the phenomenon. This framework categorizes stability into three primary states: voltage stability, frequency stability, and angle stability. In addition, it recognizes that features can be extracted from real-time waveform measurements (IEEE, 2004). To reflect modern technological developments, the committee has also proposed adding two additional categories: converter-driven stability and resonance stability\cite{IEEE2004Taxonomy}.
Within the voltage stability classification, the distinction is between short-term voltage stability (STVS) and long-term voltage stability, based on time scale. Unlike long-term voltage stability—which is governed by slower mechanisms such as load tap changers and thermostatic controls—STVS focuses on fast post-disturbance voltage dynamics occurring within approximately 10–20 seconds following a large disturbance.

This classification is widely accepted and highly effective in simplifying power system stability analysis. However, based on an extensive literature review of STVS \cite{Sarwar_Review}, it is evident that different time-scale phenomena are often grouped under the same STVS category. Following the logic of the IEEE/CIGRE framework, we propose further decomposing STVS into three subcategories: slow dynamics, fast dynamics, and very fast dynamics. This refined classification is beneficial not only for assessment but also for analysis and modeling.

Slow dynamics include electro-thermal phenomena, primarily caused by thermal relays in single-phase induction motors (in aggregated load models). These are often described as delayed voltage recovery (FIDVR), with a typical time scale of about 10 seconds. During a fault-induced voltage dip, stalled induction motors draw high reactive power, which increases winding temperature and triggers thermal relays. Depending on the thermal relay characteristics and system operating conditions—particularly the share of the induction motor load—the voltage may either recover to its nominal level or remain at a reduced value. Successful recovery may occur quickly enough to prevent additional voltage-dependent relay operations or slowly enough to trigger other components in the system. From a modeling perspective, these behaviors can be adequately captured using time-series power flow combined with thermal relay models, without requiring full dynamic simulations \cite{time_series_FIDVR}.

Fast dynamics correspond to electromechanical phenomena, in which three-phase induction motors can induce oscillations in the voltage magnitude, known as transient voltage response (TVR) in the North American Electric Reliability Corporation (NERC) \cite{NERC2017TVR}. The literature generally reports three types of behavior in this category: stable oscillations, unstable oscillations, and chaotic responses. Capturing stable and unstable oscillations requires time-domain simulation with detailed dynamic models of loads and generators in the phasor domain, while chaotic behavior necessitates careful handling of numerical errors to avoid misleading results. Based on extensive studies on the Nordic test system with three-phase induction motor loads, it is observed that voltage magnitudes across different buses oscillate with the same frequency (mode) and exhibit coherent modal shapes. This suggests that the system dynamics can be approximated by a reduced-order representation, potentially as a one-dimensional (1-D) dynamic system. However, a purely 1-D representation based only on voltage magnitude is insufficient to fully capture the oscillatory behavior. This limitation indicates that voltage magnitude alone does not provide a complete description of short-term voltage stability dynamics. To address this, an additional state variable—specifically the rate of change of voltage (ROCOV)—should be incorporated. The need for such augmentation is indirectly supported by classical reduced-order voltage stability models, such as those proposed by Ian Dobson in 1988 \cite{194705}. However, despite this insight, the explicit inclusion of ROCOV as a state variable has not been systematically considered in voltage stability assessment studies.

Very fast dynamics include fast voltage collapse phenomena, often associated with network singularities and loadability limits. Since these phenomena are strongly influenced by network behavior, electromagnetic transient (EMT) simulations are recommended rather than purely algebraic models. This category may also include intermediate phenomena (faster than electromechanical and shower than electromagnetic), such as oscillations driven by converter controls, phase-locked loops (PLLs), or switching dynamics. Due to their fast time scale, these phenomena require high-resolution measurements, such as those from disturbance fault recorders or advanced synchronized measurement technologies (e.g., Waveform Measurement Units (WMU) \cite{10820137}).

In this work, we focus on slow and fast dynamics, as they can be effectively captured using PMU measurements. A review of the literature shows that most STVS studies fall into one of these categories, but are typically analyzed independently. For example, FIDVR studies assume smooth voltage recovery without oscillations, while oscillatory studies assume fluctuations around nominal voltage without delayed recovery. These assumptions are not fully realistic. In actual power systems, both single-phase induction motors (causing delayed recovery) and three-phase induction motors (causing oscillations) coexist. Consequently, the system exhibits multi-time-scale dynamics, as also observed in PMU measurements from real systems such as the Colombian power grid \cite{10328255}.

In addition to the complexity introduced by multi-time-scale dynamics, short-term voltage stability (STVS) is inherently a high-dimensional, nonlinear dynamic problem. The presence of time-dependent components—such as thermal relays and voltage-dependent protection schemes (e.g., low/high voltage ride-through)—introduces non-autonomous behavior into the system. In contrast to transient angle stability, which primarily focuses on the dynamics of synchronous generators and often represents loads using steady-state models, STVS cannot be readily reduced to a set of ordinary differential equations (ODEs) through techniques such as Kron reduction. This is because loads in STVS studies are modeled dynamically, and the system is further complicated by ill-conditioning arising from the singularity of the algebraic equations. Moreover, STVS is typically triggered by large disturbances near load buses, leading to significant deviations from the equilibrium point. Under such conditions, linearization becomes inadequate for accurately capturing system behavior. These challenges motivate the development of a nonlinear, trajectory-based assessment index, such as one based on Lyapunov exponents, to effectively characterize system stability. From a practical perspective, fast real-time assessment of STVS is critical for preventing instability or unwanted tripping, particularly if the index can predict events associated with delayed voltage recovery. Existing literature predominantly focuses on stability assessment, with limited attention to tripping detection. However, in practice, many blackouts originate from stressed yet initially stable operating conditions that progressively evolve into cascading failures due to protection actions. The recent blackout in Spain is an example of that \cite{ICS2026SpainPortugalIncident}.


Moreover, most existing real-time STVS assessment methods provide only a binary classification of system behavior as stable or unstable. However, system operators require more informative indicators that quantify the degree of stability and the proximity to instability, particularly to support situational awareness, contingency ranking, and corrective control actions. In response to this need, prior work on STVS assessment can be broadly categorized into three methodological classes: model-based approaches, data-driven techniques, and trajectory-based methods (using LE). 

Model-based approaches rely on analytical representations of system dynamics and stability theory, typically using Lyapunov or energy functions derived from the second Lyapunov method. Early contributions by Hiskens and Davy \cite{Hiskens1996LoadEnergy} introduced energy-function formulations that incorporate generator and load dynamics to estimate voltage stability margins and critical clearing times. Subsequent work extended these ideas to transient conditions and explicitly accounted for induction motor behavior, enabling physically interpretable margins for STVS assessment \cite{Meintjes1990TEFunctionMotor, praprost2002energy}. While these methods offer strong theoretical guarantees, their reliance on simplified models and offline parameterization limits their scalability and accuracy for real-time applications, particularly in systems with diverse load compositions and high IBR penetration. Moreover, constructing suitable Lyapunov or energy functions for such heterogeneous systems remains challenging and often introduces conservativeness.

To overcome these limitations, data-driven and artificial intelligence–based methods have been widely explored for real-time STVS assessment. Early studies employed classical machine learning techniques to extract discriminative features from voltage trajectories, including shapelet-based classifiers \cite{Zhu2015Shapelet, Zhu2017Imbalance}, least-squares support vector machines \cite{Yang2018LSSVM}, and random forest models \cite{Pinzon2019RF}. Ensemble learning approaches further improved robustness under noisy and high-dimensional measurements \cite{Babaali2023Ensemble}. More recently, deep learning models have been proposed to capture the spatial–temporal dependencies inherent in voltage dynamics. Representative examples include spatial–temporal graph convolutional networks that embed system topology \cite{Luo2021STGCN} and Transformer-based transfer learning frameworks that adapt to changing operating conditions with minimal retraining \cite{Li2023Transformer}. Although these methods achieve high classification accuracy and fast response, they typically require large labeled datasets and offer limited physical interpretability, motivating the development of alternative frameworks that directly exploit system dynamics.

Trajectory-based methods grounded in Lyapunov’s first (indirect) method offer a promising compromise between physical insight and real-time applicability. These approaches evaluate post-disturbance voltage behavior directly from measurements, avoiding the need for detailed system models or extensive offline simulations. Dasgupta et al. \cite{dasgupta2013real} introduced a real-time framework for computing finite-time Lyapunov exponents (FTLEs) from PMU data, explicitly addressing practical challenges such as noise and finite observation windows while enabling bus-level stability assessment. Subsequent work refined the estimation process through offline stochastic identification to improve robustness across operating conditions \cite{Pinzon2019PMU_STVS}. By quantifying the exponential divergence or convergence of nearby trajectories, Lyapunov exponents provide a nonlinear measure of system stability: a negative maximum exponent indicates convergence, whereas a positive value signals instability \cite{LE1}. However, conventional applications of Lyapunov exponents remain largely qualitative and are sensitive to noise. Following major disturbances, the maximum Lyapunov exponent may oscillate between positive and negative values for extended periods due to the coexistence of multiple dynamic modes, complicating stability interpretation \cite{LE3}. Koopman-operator–based extensions have been proposed to accelerate detection \cite{koopman_LE1, koopman_LE2}, but their real-time implementation and operational robustness in large-scale power systems remain open challenges.

Beyond these methodological limitations, existing STVS assessment frameworks rarely quantify the stability or explicitly distinguish between recovery-driven and oscillatory dynamics. In particular, no real-time method has incorporated the dynamics of single-phase induction motors, which play a dominant role in slow voltage recovery during FIDVR events. Voltage trajectories under such conditions inherently contain two distinct dynamical modes: a slow mode associated with thermal-relay–driven induction motor behavior, producing an approximately slow exponential recovery, and a fast oscillatory mode. While single-phase motors eventually either reaccelerate or trip, this state-dependent transition is not directly observable from voltage measurements. Consequently, applying Lyapunov exponents to the composite signal may yield mixed stability indicators—such as a positive exponent associated with exponential recovery and a negative exponent associated with oscillations—even when the system is ultimately stable. From an operational perspective, the critical concern during slow recovery remains the risk of voltage-dependent protection activation rather than long-term divergence.

To address these gaps, this paper proposes a trajectory-based nonlinear framework for real-time STVS quantification that explicitly separates oscillatory and residual recovery dynamics. By combining empirical mode decomposition (EMD), Lyapunov exponent analysis, and Kullback–Leibler (KL) divergence, the proposed approach enables physically meaningful stability assessment beyond binary classification. The main contributions of this work are summarized as follows:
\begin{itemize}
    \item The proposed method separates slow and fast voltage dynamics, quantifies the proximity to instability of oscillatory behavior, and predicts relay tripping associated with delayed voltage recovery.

    \item The framework introduces a two-stage time-delay embedding reconstruction to capture ROCOV dynamics directly from voltage trajectories. This enhancement significantly improves the responsiveness of Lyapunov-based assessment, enabling stability prediction within 0.6~s after a disturbance, compared to approximately 10~s required by conventional methods.

    \item A quantitative stability index is introduced to measure the degree of short-term voltage stability, moving beyond a binary stable/unstable classification.
\end{itemize}

The remainder of this paper is organized as follows. Section~II reviews the theoretical background on Lyapunov exponents, empirical mode decomposition, and KL divergence. Sections~III and IV present the proposed stability index and threshold selection, respectively. Section~V discusses simulation results using the Nordic test system, and Section~VI concludes the paper.

\section{Background}
\subsection{Lyapunov Exponents (LEs)}
\label{subsec:LE_background}
\subsubsection{Definition} 
\begin{definition}[Lyapunov Exponent]
The \emph{Lyapunov exponent} (LE), also known as the Lyapunov characteristic exponent, quantifies the average exponential rate of divergence or convergence between two infinitesimally close trajectories of a dynamical system in phase space.  

Let $V(t)$ and $V_0(t)$ denote two trajectories with an initial deviation
\(
\delta V(t) = V(t) - V_0(t), \quad 
\delta V_0 = V(0) - V_0(0).
\)
If their separation evolves approximately as
\begin{equation}
    \|\delta V(t)\| \approx e^{\lambda t}\,\|\delta V_0\|,
    \label{LE_voltage}
\end{equation}
then $\lambda$ is defined as the \emph{Lyapunov exponent} \cite{LE_defenition}. A positive value of $\lambda$ indicates exponential divergence (instability), whereas a negative value implies exponential convergence (stability).
\end{definition}

Lyapunov exponents can be characterized along two fundamental dimensions \cite{LE_Local_global}:  
(i) \emph{global} versus \emph{local}, depending on whether they are computed over an infinite or finite time horizon, and  
(ii) \emph{linear} versus \emph{nonlinear}, depending on whether they are derived from infinitesimal or finite-size perturbations, respectively. Over a finite time window $[t_0, t_0+T]$, the FTLE is defined as
\begin{equation}
    \lambda_{\text{FTLE}}(t_0,T)
    = \frac{1}{T}\ln\frac{\|\delta v(t_0+T)\|}{\|\delta v(t_0)\|}.
    \label{eq:FTLE}
\end{equation}
When perturbations are infinitesimal, \eqref{eq:FTLE} is computed from the linearized variational equations in the tangent space, corresponding to the finite-time linear LE \cite{FTLE}. For perturbations of finite magnitude, the measure generalizes to the \emph{finite-size Lyapunov exponent (FSLE)}, which evaluates the time required for an initial deviation to grow by a prescribed amplification factor \cite{NLFTLE,FSLE}.

The existence of Lyapunov exponents is formally established under ergodic conditions. In one-dimensional systems, their existence follows from the Birkhoff ergodic theorem \cite{yosida1939birkhoff}, while in higher-dimensional settings, the multiplicative ergodic theorem by Oseledec \cite{oseledec1968multiplicative} guarantees the existence of a complete spectrum of Lyapunov exponents. From a linear LE perspective, perturbation dynamics evolve according to a fundamental state transition matrix, whose singular values determine the finite-time growth rates; the logarithms of these rates approximate the FTLEs. For hyperbolic equilibria, LEs correspond to the real parts of system eigenvalues, whereas for periodic orbits, they coincide with Floquet exponents \cite{vulpiani2010chaos}. According to the Oseledec splitting theorem, the state space can be decomposed into nested invariant subspaces, each associated with a specific LE; random perturbations typically align with the most unstable subspace, corresponding to the largest exponent \cite{LE_book}. The sign of the maximal exponent $\lambda_1$ serves as a key indicator of system behavior: negative values signify convergence and stability, zero indicates neutral or marginal stability (as in limit cycles), and positive values denote exponential divergence and sensitivity to initial conditions—a defining feature of deterministic chaos. Multiple positive exponents imply hyperchaotic behavior. 

In power-system applications, practical computation of finite-time or finite-size LEs requires careful selection of the observation window to capture transient post-fault behavior relevant to STVS. These methods remain robust even for piecewise-differentiable dynamics or events involving discontinuities, such as protection actions, since the focus lies on trajectory separation rather than smoothness. In practice, the magnitude and sign of the MLE serve as early indicators of dynamic instability, yet should be complemented by signal decomposition techniques—such as distinguishing residual and oscillatory components—to avoid conflating delayed voltage recovery with undamped oscillations.

A measurement-based approach for estimating the FTLE was introduced in \cite{LE2} and further developed in \cite{dasgupta2013real}, where the exponent is computed directly from voltage measurements over a finite data window. This method enables real-time estimation of local divergence rates without requiring explicit system models, making it suitable for PMU-based stability monitoring. However, as noted in \cite{LE3}, the approach is highly sensitive to oscillations in the voltage trajectory, which can distort the inferred stability condition. In particular, when the voltage exhibits damped oscillations combined with exponential recovery, the instantaneous slope of the trajectory may temporarily increase even under stable conditions, leading the algorithm to falsely indicate instability. Such oscillation-induced fluctuations highlight a key limitation of measurement-based LE estimation, underscoring the need for enhanced techniques that can distinguish genuine divergence from transient oscillatory effects in short-term voltage stability analysis.

\subsubsection{Impact of Measurement Noise on FTLE Estimation}

In practical applications, the finite-time Lyapunov exponent (FTLE) is estimated from measured trajectories, which are inevitably corrupted by noise. Even when the measurement noise has zero mean, the nonlinear logarithmic operation in the FTLE definition introduces both a bias in the mean estimate and a nonzero estimation variance. As a result, reliable FTLE estimation requires sufficiently long observation windows.

Consider the FTLE computed from the separation between two nearby trajectories:
\begin{align}
\hat{\lambda}(T) = \frac{1}{T}\ln\left(\frac{\|\delta(T)\|}{\|\delta(0)\|}\right),
\end{align}
where $\delta(t)$ denotes the deviation vector. Assume that the measured deviation at time $T$ is corrupted by additive noise:
\begin{align}
\|\delta(T)\|_{\mathrm{m}} = \|\delta(T)\| + \eta,
\end{align}
where $\eta \sim \mathcal{N}(0,\sigma^2)$. The noisy FTLE estimate becomes
\begin{align}
\hat{\lambda}_{\eta}(T)
= \frac{1}{T}\ln\left(\frac{\|\delta(T)\|+\eta}{\|\delta(0)\|}\right).
\end{align}
Equivalently,
\begin{align}
\hat{\lambda}_{\eta}(T)
= \frac{1}{T}\ln\left(\frac{\|\delta(T)\|}{\|\delta(0)\|}\right)
+ \frac{1}{T}\ln\left(1+\frac{\eta}{\|\delta(T)\|}\right).
\end{align}
Hence,
\begin{align}
\hat{\lambda}_{\eta}(T)
= \hat{\lambda}(T) + \frac{1}{T}\ln\left(1+\frac{\eta}{\|\delta(T)\|}\right).
\end{align}

When the noise is small relative to the signal, i.e., \(
\left|\frac{\eta}{\|\delta(T)\|}\right| \ll 1,
\)
the logarithm can be approximated using the Taylor expansion \(
\ln(1+x) \approx x - \frac{x^2}{2}.
\)
Substituting $x = \eta/\|\delta(T)\|$  and taking expectation yields
\begin{align}
\mathbb{E}[\hat{\lambda}_{\eta}(T)]
\approx \hat{\lambda}(T)
+ \frac{1}{T}\left(
\frac{\mathbb{E}[\eta]}{\|\delta(T)\|}
-\frac{1}{2}\frac{\mathbb{E}[\eta^2]}{\|\delta(T)\|^2}
\right).
\end{align}
Since $\eta$ has zero mean and variance $\sigma^2$, we obtain
\begin{align}
\mathbb{E}[\hat{\lambda}_{\eta}(T)]
\approx \hat{\lambda}(T)
-\frac{1}{2T}\frac{\sigma^2}{\|\delta(T)\|^2}.
\end{align}

Therefore, even though the measurement noise is zero mean, the FTLE estimate experiences a negative mean shift. Next, the estimation variance can be approximated from the first-order term:
\begin{align}
\mathrm{Var}\big(\hat{\lambda}_{\eta}(T)\big)
\approx
\frac{1}{T^2}\frac{\mathrm{Var}(\eta)}{\|\delta(T)\|^2}
=
\frac{\sigma^2}{T^2\|\delta(T)\|^2}.
\end{align}

This result shows that the variance decays proportionally to $1/T^2$. Consequently, short observation windows produce high-variance FTLE estimates, while longer windows are required to suppress the effect of measurement noise. In other words, even for a stable trajectory, measurement noise can distort the FTLE estimate through both:
\begin{itemize}
    \item a \textbf{mean shift} caused by the nonlinear logarithmic transformation, and
    \item a \textbf{variance term} that decreases as $1/T^2$.
\end{itemize}

\subsubsection{Impact of Two-Time-Scale Dynamics on FTLE Estimation}

To illustrate the impact of two-time-scale behavior on FTLE calcaulation, consider the following linear two-time-scale system:
\begin{align}
\dot{x} &= -a x + \omega y + b z, \\
\dot{y} &= -\omega x - a y, \\
\dot{z} &= -\epsilon z,
\end{align}
where $\omega \gg a > 0$, $0 < \epsilon \ll a$, and $b \neq 0$. The states $(x,y)$ represent fast oscillatory dynamics, while $z$ captures the slow mode. The system is exponentially stable.  Defining the complex variable $w = x + j y$, the system can be rewritten as:
\begin{align}
\dot{w} = (-a - j\omega)w + b z, \quad z(t) = z_0 e^{-\epsilon t}.
\end{align}

Solving for $w(t)$ yields:
\begin{align}
w(t) = b z_0 \frac{e^{-\epsilon t} - e^{-(a + j\omega)t}}{a + j\omega - \epsilon}.
\end{align}

Within the intermediate time window:\(
\frac{1}{a} \ll t \ll \frac{1}{\epsilon},
\)
the fast transient has already decayed ($e^{-a t} \approx 0$), while the slow mode remains nearly constant ($e^{-\epsilon t} \approx 1$). 

\begin{align}
|w(t)| \approx |z_0| e^{-\epsilon t} \frac{|b|}{\sqrt{(a - \epsilon)^2 + \omega^2}}, \quad |z(t)| = |z_0| e^{-\epsilon t}.
\end{align}

Thus, the overall trajectory norm behaves as:
\begin{align}
\|(x(t), y(t), z(t))\| \approx |z_0| e^{-\epsilon t} 
\sqrt{1 + \frac{b^2}{(a - \epsilon)^2 + \omega^2}}.
\end{align}

So, the FTLE is: 
\begin{align}
\lambda(T) = -\epsilon + \frac{1}{2T} \ln \sqrt{1 + \frac{b^2}{(a - \epsilon)^2 + \omega^2}}
\end{align}

Even when the system is dynamically stable, the estimated FTLE can become positive for short observation windows, specifically when
\(
T < \frac{1}{4\epsilon} \ln \left(1 + \frac{b^2}{(a - \epsilon)^2 + \omega^2} \right),
\)
which may lead to misleading stability assessment. By varying the coefficients of the previously introduced two-time-scale system, the impact of fast and slow dynamics on FTLE convergence time can be systematically analyzed. The results are summarized in Table~\ref{Table1}. 

In general, the FTLE convergence time can be categorized as follows: short convergence occurs on the order of $T \sim \mathcal{O}(1/\lambda_{\text{fast}})$, medium convergence requires a few multiples of this time scale, long convergence is governed by the slow dynamics with $T \sim \mathcal{O}(1/\epsilon)$, and very long convergence corresponds to cases where $T \gg \mathcal{O}(1/\epsilon)$ due to strong interaction between modes. In non-normal systems, different stable modes can temporarily reinforce each other before decaying, leading to delayed FTLE convergence. These results explain why conventional FTLE-based methods struggle to provide fast and reliable stability assessment in STVS problems.

\begin{table}[!t]
\centering
\caption{Impact of two-time-scale dynamics on FTLE convergence time}
\label{Table1}
\begin{tabularx}{\columnwidth}{|X|X|X|}
\hline
\textbf{Fast Dynamics} & \textbf{Slow Dynamics} & \textbf{FTLE Convergence Time $T$} \\
\hline
Real, monotone decay & Any & Short \\
\hline
Oscillatory decay & Relatively fast recovery & Medium \\
\hline
Oscillatory decay & Slow recovery & Long \\
\hline
Strong non-normal transient growth & Relatively fast recovery & Medium \\
\hline
Strong non-normal transient growth & Slow recovery & Very Long \\
\hline
\end{tabularx}
\end{table}

\subsection{Empirical Mode Decomposition (EMD)}

EMD is a powerful, adaptive signal processing technique introduced by N. E. Huang and colleagues in 1998 \cite{EMD8}. It is specifically designed to analyze non-linear and non-stationary time series data by decomposing the original signal into a set of Intrinsic Mode Functions (IMFs) and a residual (R). Each IMF represents a simple oscillatory mode (i.e., the number of peaks equals the number of zero crossings plus 1), capturing different frequency components of the signal. This process is analogous to an adaptive wavelet transform but without requiring predefined basis functions. The flexibility, computational process and data-driven nature of EMD make it particularly suitable for our real-time application \cite{wang2010intrinsic}.

The Multivariate Empirical Mode Decomposition (MEMD) \cite{6633076} is employed to extract a set of intrinsic mode functions (IMFs) that jointly represent the oscillatory components across all signal channels. 
Each IMF captures a distinct oscillatory mode that is aligned across variables, ensuring consistent mode representation among multivariate data. 
The final residue \(R(t)\) represents the slowly varying trend common to all channels. 
Mathematically, for each signal \(v_m(t)\) in the multivariate dataset, the MEMD-based decomposition can be expressed as:
\begin{equation}
v_m(t) = \sum_i \text{IMF}_{i,m}(t) + R_m(t)
\end{equation}
where \(\text{IMF}_{i,m}(t)\) denotes the \(i^{\text{th}}\) intrinsic mode function associated with the \(m^{\text{th}}\) signal, and \(R_m(t)\) is its corresponding residual component. 
The summation of all IMFs, \(\sum_i \text{IMF}_{i,m}(t)\), thus represents the reconstructed signal obtained from MEMD, excluding the residual trend. Beacause these IMFs are simple oscillation modes, unwanted frequency components can be readily removed by applying zero-crossing–based frequency estimation, which enables straightforward identification and filtering of undesired oscillatory content. This approach can also serve as an effective and simple noise-reduction method.

\subsection{KL Divergence Measure }

KL divergence, also known as relative entropy, is a fundamental concept in information theory used to quantify the difference between two probability distributions, \(P\) and \(Q\). Mathematically, KL divergence is defined as:
\begin{equation}
\centering
   D_{KL}(P||Q) = \sum_i P(i) \ln\frac{P(i)}{Q(i)}
\label{eq:KL} 
\end{equation}
for discrete distributions and as an integral for continuous distributions. This measure is non-symmetric and always non-negative, with \(D_{KL}(P||Q) = 0\) if and only if \(P\) and \(Q\) are identical. In the context of FIDVR, KL divergence is used to quantify the deviation of post-fault voltage profiles from the ideal behavior, denoted as \(Q=P^{ref}\), providing a rigorous measure for detecting and analyzing FIDVR events \cite{KL1}. So, it is applied to find the statistical ``distance" between the observed voltage signal and the reference ideal signal \(P^{ref}\). 
In this paper, the KL divergence is employed to quantify the statistical distance between the divergence factor (the exponential of the Lyapunov exponent) and a reversed shifted Gompertz distribution, thereby providing an indirect measure of how far the Lyapunov exponent deviates from the zero axis.

\section{Proposed Method}

The proposed framework begins by decomposing the post-fault voltage trajectory using MEMD into two distinct components:
The residual component captures the slow recovery trend and identifies delayed recovery events, and the Oscillatory component, represented by IMFs, captures oscillatory behavior. For the residual component, the FSLE is computed for each individual generator voltage profile to evaluate the speed of recovery. For the oscillatory component, FTLE is calculated at the system level to measure how voltage trajectories converge or diverge, thereby capturing the oscillatory form of stability. The resulting distributions of the exponential of these exponents (which are called divergence factors) are then compared with a reference profile using the KL divergence, yielding two complementary indices: recovery stability index and oscillation stability index.

\begin{figure*}
    \centering
    \includegraphics[width=\textwidth]{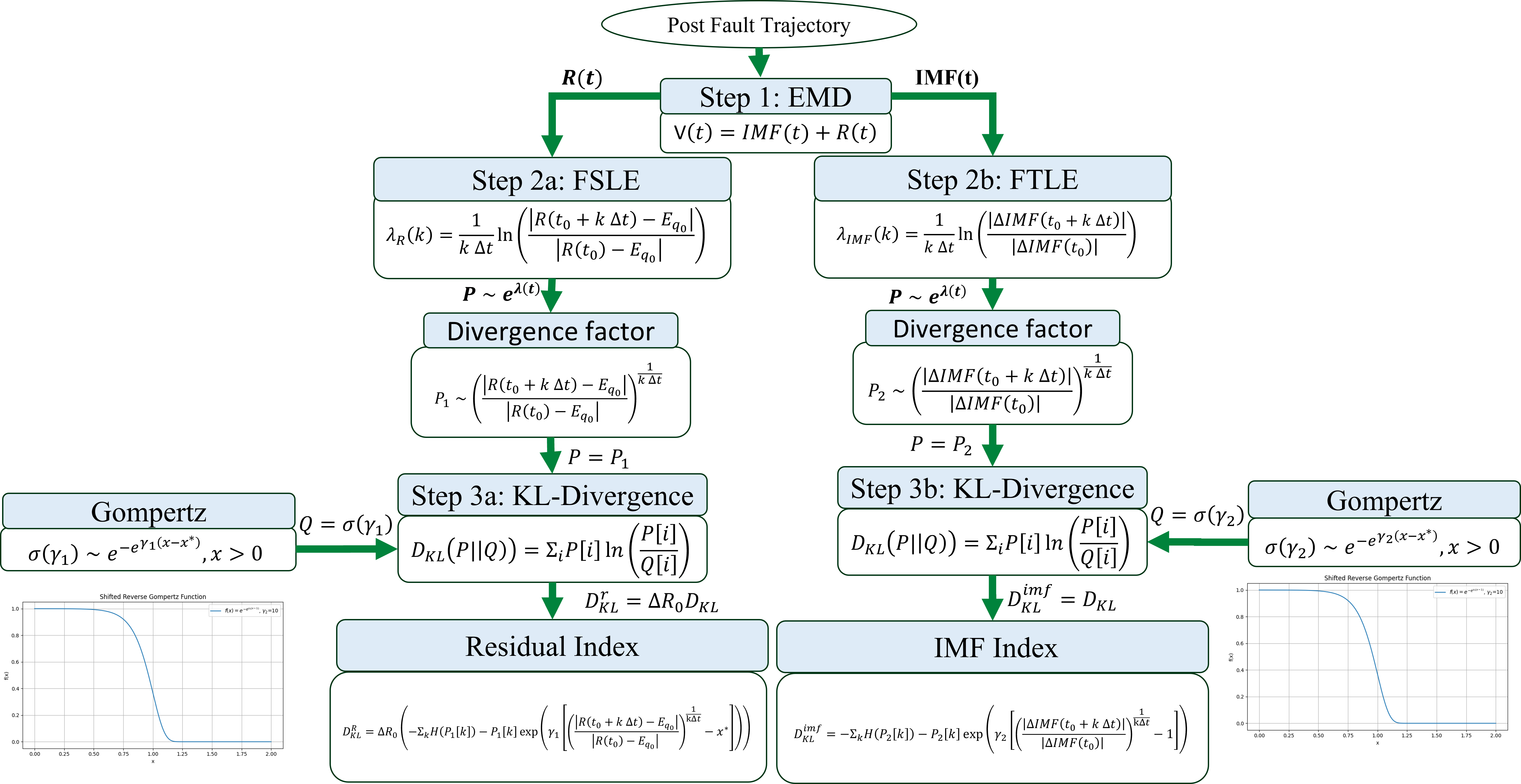}
    \caption{Flowchart of the proposed Short-Term Voltage Stability Index (STVSI). The procedure begins with Empirical Mode Decomposition (EMD) of the post-fault trajectory, followed by Lyapunov Exponent (LE) computation, probability distribution construction, KL-divergence evaluation, and final index calculation.}
    \label{fig:flowchart}
\end{figure*}

By construction, these indices increase as the system approaches instability, thereby enabling both classification and measurement of the proximity to the critical operating boundary. Critical threshold values are defined to separate stable, unstable, and near-critical cases. A high-level flowchart of the proposed STVSI framework is shown in Fig.~\ref{fig:flowchart}.

\subsection{Recovery Stability Index}

The first component of the proposed framework addresses the impact of delayed voltage recovery, which can initiate cascading failures through activation of over-excitation limiters (OEL) in synchronous generators or low-voltage ride-through relays (LVRT) in inverter-based resources. Within this framework, the proposed index predicts potential tripping of synchronous generators caused by delayed voltage recovery. After the first step, extracting the residual compoenent using EMD, the FSLE is computed for each generator voltage profile individually as shown in step 2a in Figure \ref{fig:flowchart}. In practice, large disturbances can drive the system trajectory far from both the pre- and post-fault equilibrium points. Therefore, FSLE is applied here to quantify the rate at which the trajectory reconverges toward the pre-fault equilibrium point. In this formulation, the equilibrium point itself serves as the reference trajectory, and the FSLE is expressed as:

\begin{equation}
\lambda_{R}(k) = \frac{1}{k \Delta t} \ln \left( \frac{| R(t_0 +K \Delta t) - Eq_{0} |}{| R(t_0 ) - Eq_{0} |} \right),
\label{eq:fsle_residual}
\end{equation}
where $Eq_{0}$ denotes the post-fault equilibrium voltage, $\Delta t$ is the PMU sample rate. A negative value of $\lambda_{R}(k)$ indicates convergence of the residual trajectory toward the equilibrium, while a positive value indicates divergence. To transform the FSLE values into an interpretable statistical measure, we construct the probability distribution $P_{1}$ of the exponential of$\lambda_R$ which represent the divergence factor:
\begin{equation}
P_{1} \sim \exp \left( \lambda_{R}(t) \right)=\left( \frac{| R(t_0 +k\Delta t) - Eq_{0} |}{| R(t_0) - Eq_{0} |} \right)^{\frac{1}{k \Delta t}}.
\label{eq:prob_residual}
\end{equation}
The range of $\lambda_{R}$ for converging trajectories is $(-\infty, 0]$, while for diverging trajectories it is $\lambda_{R} > 0$. However, for divergance factor, the range of $\exp(\lambda_{R})$ is mapped to $(0,1]$ for convergent cases and to values greater than $1$ for divergent cases. The deviation of $P_{1}$ from a reference signal is quantified using the KL divergence in step 3a, where the reference is modeled by a shifted-reversed Gompertz distribution that effectively represents asymptotic convergence toward equilibrium while penalizing the slow recovery. The mathematical representation of shifted-reversed Gompertz is given by: 

\begin{equation}
\sigma(x) = e^{-e^{\,\gamma_1 (x-x^*)}}, \quad x > 0,
\label{eq:gomp_standard}
\end{equation}
where $\gamma_1$ is a shape parameter that controls the steepness of the distribution, and $x^*$ is the shifted value. Because the speed of recovery alone is insufficient to predict the triggering of OEL or LVRT, the index is weighted by the depth of the initial voltage dip. Specifically, $\Delta R_0=|V_{pre}-R_0|$ is introduced as a multiplicative weight, where $R_{0} = R(k_0 \Delta t)$ denotes the initial residual voltage after the fault and $V_{pre}$ is the voltage value before the fault. This weighting ensures that shallow dips, even with slow recovery, are not overemphasized, while deeper dips with slow recovery are correctly flagged as dangerous situations. With these considerations, the recovery stability index is defined as:
\begin{equation}
D^{R}_{KL} = \Delta R_{0} \times D_{KL} \left( P_{1} \,\|\, \sigma(x;\gamma_{1},x^*) \right),
\label{eq:kl_residual}
\end{equation}
 The KL divergence itself is given by \ref{eq:KL}. An explicit, expanded formulation of the recovery stability index is given in (\ref{eq:kl_residual_expanded}), which is presented at the bottom of the this page.
where $\mathbf{H}(x) = -x \ln(x)$ is the entropy function.  This form highlights the dependence of the index on both the convergence speed of the residual trajectory toward the post-fault equilibrium $Eq_{0}$ and the depth of the initial voltage dip $R(t_{0})$. It also makes explicit the role of the tuning parameters $\gamma_{1}\text{ and } x^*$ in adjusting the sensitivity of the measure.

\begin{figure*}[!b]
\centering
\setlength{\abovedisplayskip}{4pt}
\setlength{\belowdisplayskip}{4pt}
\begin{minipage}{0.96\textwidth}
\begin{equation}
D_{\text{KL}}^R = \Delta R_0\;
\Bigg[
- \sum_{k} \mathbf{H}\!\left(P_{1}[x_k]\right)
- P_{1}[x_k] \, \exp\!\Bigg(
\gamma_{1} \Bigg[
\left(\frac{|R(k\Delta t+t_0)-Eq_0|}{|R(t_0)-Eq_0|}\right)^{\tfrac{1}{k\Delta t}} - x^* 
\Bigg]\Bigg)
\Bigg],
\label{eq:kl_residual_expanded}
\end{equation}
\end{minipage}
\vspace{-0.35\baselineskip} 
\end{figure*}

A smaller value of $D^{R}_{KL}$ indicates that the recovery trajectory closely follows the reference (fast recovery and stable behavior), while a larger value signifies deviation toward slow recovery and potential triggering low voltage dependent relay. By setting a critical threshold $D^{R}_{\text{critical}}$, discussed in the following section, the index can classify operating conditions as:
\begin{itemize}
    \item \textbf{Non trip:} $D^{R}_{KL} < D^{R}_{\text{critical}}$,
    \item \textbf{Trip:} $D^{R}_{KL} > D^{R}_{\text{critical}}$,
    \item \textbf{Margin:} $(D^{R}_{KL} - D^{R}_{\text{critical}})/D^{R}_{\text{critical}}*100\%$
\end{itemize}

The ``margin'' provides a quantitative measure of proximity to tripping. Importantly, the index is a \emph{one-to-one but nonlinear} mapping. This means that if the margin for generator $X$ is $-20 \%$ and for generator $Y$ is $-40\%$, we can guarantee that generator $X$ will trip before generator $Y$ if the operating condition continues to move in the same direction. However, the increment required to move the margin from $20\%$ to $30\%$ is not necessarily equal to the increment required to move it from $30\%$ to $40\%$. In other words, the Recovery Stability Index is smooth and monotonic, ensuring consistent ordering of risk across generators, but it does not imply linear scaling with respect to system changes.

\subsection{Oscillation Stability Index}

The second component of the proposed framework addresses instability driven by undamped oscillations. The oscillation instability manifests itself through sustained or growing oscillations around the nominal voltage. To capture this behavior, we analyze the oscillatory components of the post-fault voltage trajectory, represented by the IMFs. As illustrated in step 1 in Figure \ref{fig:flowchart}. Then, FTLE is computed at the system level for all IMFs.  If we represent the vector of $IMF(t) = [\sum_i IMF_{i,1} (t),\quad  \sum_i IMF_{i,2} (t) , .. \sum_i IMF_{i,n}]^T $ then FTLE is expressed as:
\begin{equation}
\lambda_{\text{IMF}}(k) = \frac{1}{k \Delta t} \ln \left( \frac{|\delta  IMF(t_0+k \Delta t)|}{|\delta{\text{IMF}}(t_0)|} \right),
\label{eq:fsle_imf}
\end{equation}
where $|.|$ denotes the norm of the vector, and $\delta IMF(t)$ is a perpetration in IMF vector at time t. A negative value of $\lambda_{\text{IMF}}$ indicates that oscillations are damped, while positive values reflect growth of oscillations and potential instability. The divergance factor $exp(\lambda_{\text{IMF}})$ is used to construct a probability distribution $P_{2}$:
\begin{equation}
P_{2} \sim \exp\!\left( \lambda_{\text{IMF}}(t) \right),
\label{eq:prob_imf}
\end{equation}
Similar to the recovery case, stable oscillations map to the range $(0,1]$, while growing oscillations correspond to values greater than $1$. The oscillation index is obtained by measuring the KL divergence between $P_{2}$ and the shifted–reversed Gompertz distribution $\sigma(\gamma_{2})$. Formally,
\begin{equation}
D^{\mathrm{IMF}}_{KL} = D_{KL}\!\left(P_{2}\,\|\,\sigma(\gamma_{2})\right),
\label{eq:kl_imf}
\end{equation}

\begin{figure*}[!b]
\centering
\begin{equation}
D^{\mathrm{IMF}}_{KL}
= -\sum_{i} \mathbf{H}\!\left(P_{2}[x_i]\right)
 - P_{2}[x_i]\,
 \exp\!\left\{
 \gamma_{2}\!\left[
 \left(\frac{|\Delta \mathrm{IMF}(i\Delta t + t_{0})|}{|\Delta \mathrm{IMF}(t_{0})|}\right)^{\tfrac{1}{i\Delta t}}
 - 1
 \right]\right\},
\label{eq:kl_imf_expanded}
\end{equation}
\end{figure*}

For completeness, the expanded formulation of the oscillation stability index
is given in (\ref{eq:kl_imf_expanded}), which parallels the recovery case. The numerical implementation of the proposed index builds upon a modified LE estimation framework that extends the classical Rosenstein-style approach \cite{rosenstein1993practical, dasgupta2013real}. Rather than collapsing divergence dynamics into a single LE by averaging, we built the index based on time-resolved exponents $\lambda(i)$, as discussed before. Although the IMFs extracted from multi-bus voltages form an $ N$-dimensional multivariate trajectory, these raw temporal sequences do not fully capture the underlying nonlinear dynamics. Time-delay embedding enhances this representation by reconstructing the trajectory in a higher-dimensional phase space, thereby unfolding its structure and revealing the system’s underlying geometry \cite{kantz2003nonlinear}. To capture the dynamics of the rate of change of voltage (ROCOV), a two-step embedding reconstruction is proposed. In the first step, the measurement vector is augmented by incorporating a discrete derivative of the voltage signal, defined as $v_i - v_{i-1}$. This results in the reconstructed state vector:
\begin{align}
x_i = [v_i,\; v_i - v_{i-1}].
\end{align}

In the second step, a time-delay embedding is applied to the augmented state, yielding:
\begin{align}
\mathbf{y}_i = [x_i,\, x_{i+\tau},\, \ldots,\, x_{i+(m-1)\tau}] \in \mathbb{R}^{mN}.
\end{align}

where $m$ is the embedding dimension, $\tau$ is the delay, and $N$ is the number of signals included (two times the number of voltage measurements). The embedding dimension $m$ determines how many delayed copies of the $N$-dimensional vector are stacked together; if $m$ is too small, the attractor remains folded and divergence cannot be distinguished, while if $m$ is too large, the reconstructed trajectory becomes noisy and computationally heavy. The delay $\tau$ controls how far apart the stacked coordinates are taken: very small $\tau$ produces highly correlated coordinates that add little information, while very large $\tau$ breaks continuity by treating unrelated points as neighbors. In practice, $\tau$ is chosen based on statistical criteria such as the first minimum of the mutual information function. To avoid trivial correlations from temporally adjacent points, the nearest neighbor search also applies a Theiler window $\Theta$, which enforces $|i-j|>\Theta$ and ensures dynamically independent divergence tracking \cite{kantz2003nonlinear}.
 This condition prevents trivial matches between points that are temporally close but dynamically correlated, ensuring that divergence is tracked between independent trajectories.
Once embedding and neighbor selection are defined, divergence is measured through the evolution of inter-trajectory distances \(
    d_{i,j}(k) = \|\mathbf{y}_{i+k}-\mathbf{y}_{j+k}\|.
\) The LE is then obtained as the slope of the logarithmic divergence $\ln d_{i,j}(k)$ with respect to time. A smaller value of $D^{\text{IMF}}_{KL}$ indicates well-damped oscillations and stable post-fault behavior, while larger values indicate poor damping and possible divergence. Similar to the Recovery index, the critical value and index margin can be defined to measure the proximity to the instability point.

\section{Thresholds and Tuning of the Proposed Indices}

The effectiveness of the proposed indices depends on properly defined thresholds that separate stable from unstable cases.

\subsection{Oscillation Stability Index Threshold ($D_{\text{critical}}^{\text{imf}}$)}

The oscillation stability index threshold represents the KL divergence between the divergence factor distribution of a fixed-magnitude oscillation around nominal voltage and the shifted–reversed Gompertz distribution. Analytically, this threshold can be determined by evaluating a representative oscillatory signal whose divergence factor distribution is symmetrically centered and uniformly distributed across three bins centered at $x=1$. After normalizing the Gompertz distribution within the operational range, the resulting threshold becomes dependent on the number of bins used to discretize the distribution. For instance, when the distribution is divided into twenty bins over the interval  $[0.0,1.5]$ with $\gamma_2=10$, the corresponding critical value is  $D_{\text{critical}}^{\text{imf}}=2.09$. Note that the sensitivity of the threshold is controlled by $\gamma_2$, which acts as a tuning parameter: $\gamma_2 \rightarrow \infty$ yields a binary classifier, while a small $\gamma_2$ reduces accuracy. 

\subsection{Recovery Stability Index Threshold ($D_{\text{critical}}^{r}$)}

Recovery-induced instability is typically linked to the Over-Excitation Limiter (OEL) in synchronous machines or LVRT relays in IBRs. This subsection outlines the derivation of the OEL voltage characteristic and the procedure for tuning $\gamma_1$ and $x^*$ to find the critical threshold value, ($D_{\text{critical}}^{r}$).

The OEL constrains the maximum field current or, equivalently, the transient EMF magnitude $|E'|$. To relate this protection to terminal voltage ($V$), we derive the mapping between $E'$ and $V$. Neglecting armature resistance and setting the terminal voltage phasor as reference, we have
\begin{equation}
|E'|^2 = \left(V + \frac{X'_d Q}{V}\right)^2 + \left(\frac{X'_d P}{V}\right)^2.
\label{eq:ev_relation}
\end{equation}

where $P$ and $Q$ are the active and reactive power outputs, and $X'_d$ is the $d$-axis transient reactance. During stressed conditions, $Q$ varies with $V$. We estimate the $Q$–$V$ relation by \( V = K_1 Q + K_2 \) where $K_1$ and $K_2$ are estimated from early post-fault measurements. Substituting $Q$ in (\ref{eq:ev_relation}) allow us to estimate the limiting voltage $V_{\text{cap},i}$ at each OEL pickup level $E_i$,  by solving
\begin{equation} 
E_i^2 = \left(V_{\text{cap},i} + \frac{X'_d}{K_1}\frac{V_{\text{cap},i} - K_2}{V_{\text{cap},i}}\right)^2 + \left(\frac{X'_d P}{V_{\text{cap},i}}\right)^2,
\label{eq:EV_estimation}
\end{equation}

which reduces to a quartic in $V$. The solutions $\{E_i,V_{\text{cap},i}\}$ form the OEL voltage characteristic. In the case of LVRT relays for inverter-based resources, the terminal voltage–time tripping characteristics are already specified by grid codes, and therefore such prediction is not required.

The parameters $\gamma_1$ and $x^*$ are tuned by comparing residual signals at the verge of tripping. For this, we artificially create $s_1$ and $s_2$ signals located at the edge of tripping. Signal $s_1$ represents the slowest critical recovery, while $s_2$ is the fastest critical recovery that still activates OEL/LVRT. 
These signals are extracted from the LE–based trajectory prediction by constructing admissible envelopes that bound the future voltage evolution following a disturbance. Starting from short-term post-fault measurements (e.g, 3 seconds), finite-time Lyapunov exponents are used to characterize the local divergence or convergence rate of nearby trajectories, yielding a family of admissible voltage evolutions consistent with the observed dynamics. These admissible trajectories are then shifted to form these two signals. Then, the tuning is given by:
\[
\min_{\gamma_1,x^*}\;\gamma_1 \quad \text{s.t.} \quad 
\big|D_{\mathrm{KL},s_1}^{r}(\gamma_1,x^*)-D_{\mathrm{KL},s_2}^{r}(\gamma_1,x^*)\big| \leq f^{*}+\epsilon,
\]\[
f^{*} = \min_{\gamma_1,x^*}\big|D_{\mathrm{KL},s_1}^{r}-D_{\mathrm{KL},s_2}^{r}\big|.
\]


Here, $\epsilon$ denotes the detection tolerance. If $D_{KL}^r$ exceeds the threshold by more than $\tfrac{\epsilon}{2}$, the case is predicted to trip; if it is less than the threshold by $\tfrac{\epsilon}{2}$, the case is predicted to ride through safely. If the difference lies within $\tfrac{\epsilon}{2}$, the case is labeled as critical or near-miss. The recovery threshold is therefore defined as \(
D_{\text{critical}}^r = \tfrac{1}{2}\left(D_{\text{KL},s_1}^r + D_{\text{KL},s_2}^r\right).
\)
\begin{figure*}[t]
\centering
\includegraphics[width=\linewidth]{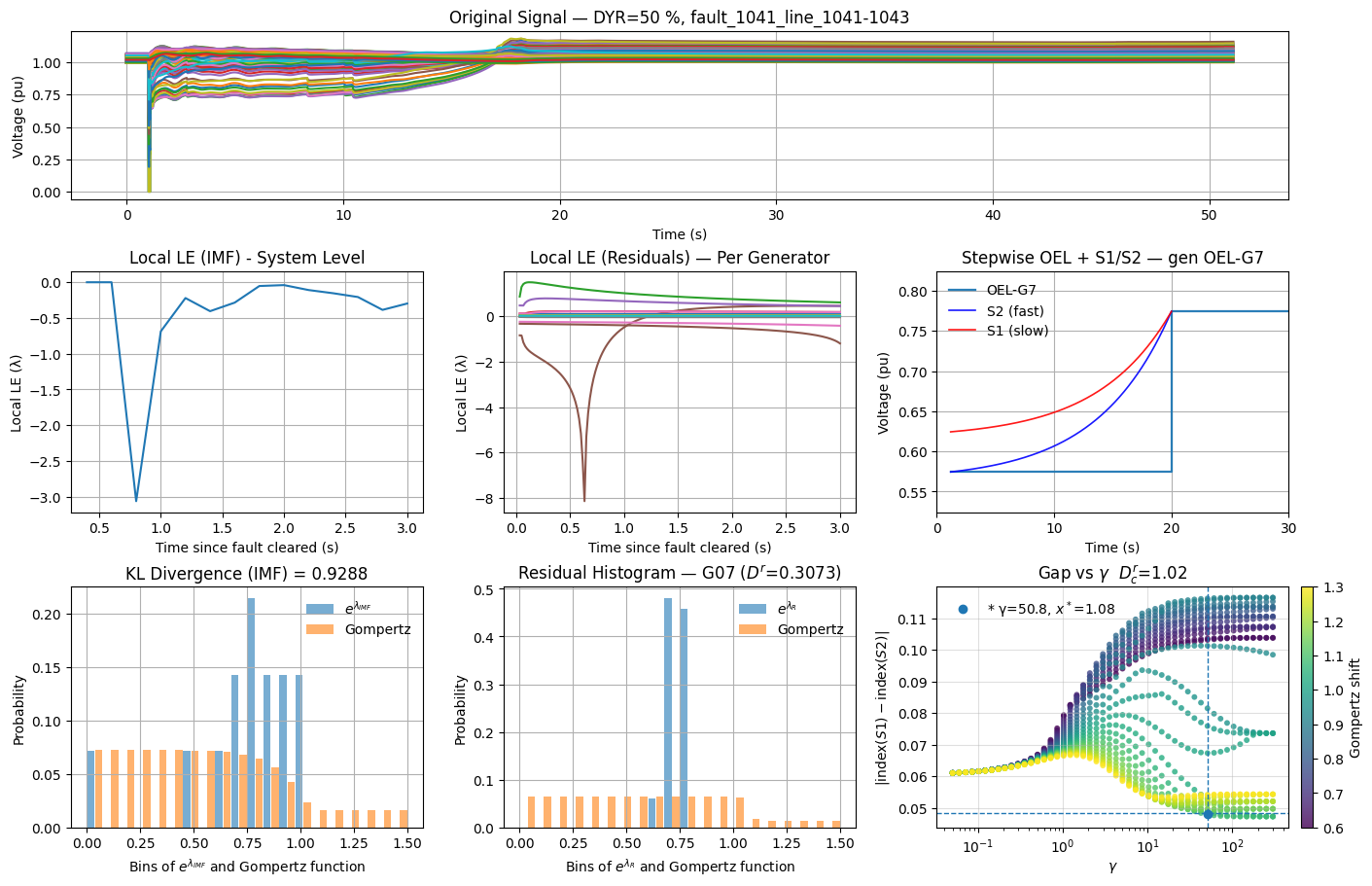}
\caption{Stability assessment for a stable case (50\% dynamic load, 
Case DA, fault at bus 1041, line 1041--1043).}
\label{fig:stable_case}
\end{figure*}
\begin{figure*}[t]
\centering
\includegraphics[width=\linewidth]{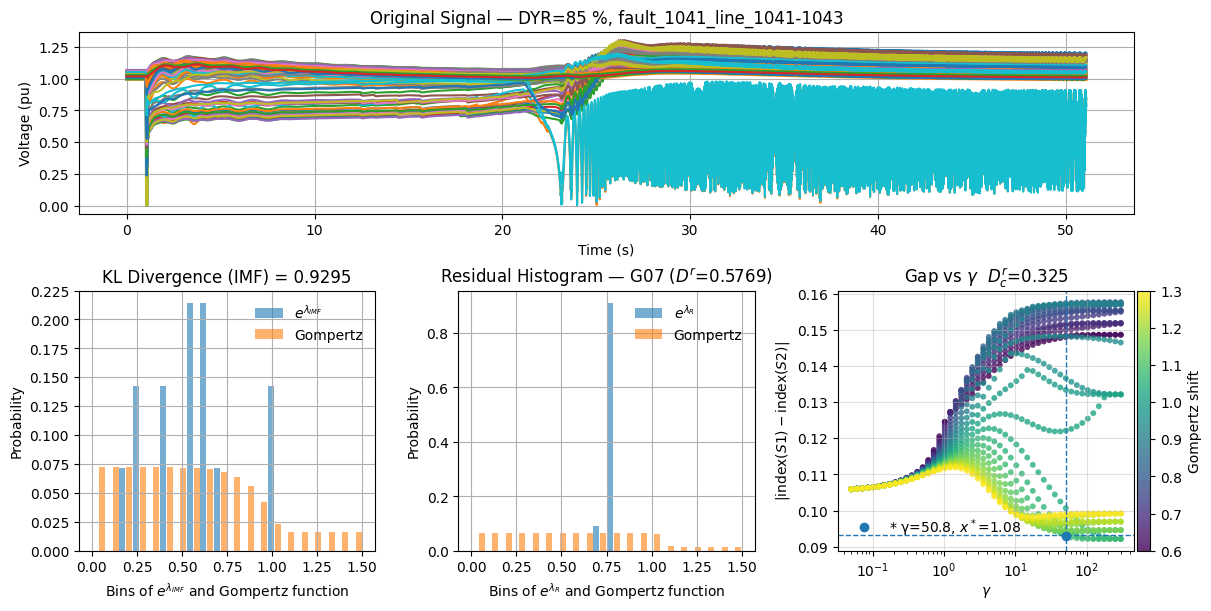}
\caption{Stability assessment for a recovery-induced unstable case 
(85\% dynamic load, Case DC, fault at bus 1041, line 1041--1043).}
\label{fig:recovery_case}
\end{figure*}

\section{Simulation and Test System Results}
\subsection{Test System Description}

The simulation was carried out using the Nordic test system at operating point~A \cite{testSystems}, 
with OELs modeled for all generators. 
Dynamic loads were represented using the Composite Load Model \cite{WECC-CMPLDW-2015}, implemented at buses 1, 2, 3, 4, 5, 41, 42, 43, 46, 47, and 51. 
Different combinations of dynamic motors are represented in the composite load model, including three-phase induction motors with varying inertia and torque characteristics (Motors A, B, C) and a single-phase induction motor capturing residential air-conditioning behavior (Motor D) 
The model parameters are adopted from the NERC recommended technical report \cite{NERC-DynamicLoad-2016}. Fault scenarios are designed by applying a three-phase fault of 100~ms duration, followed by tripping of the associated line.

\subsection{Case Study Results}

This section presents a step-by-step analysis of two representative scenarios. 
The workflow begins with the original voltage trajectories, then extracts the 
post-fault signal, applies EMD, computes local LEs, evaluates the probability distributions, and finally derives the stability indices using KL divergence.

Figure~\ref{fig:stable_case} illustrates the analysis results for the stable case corresponding to a fault on line~1041--1043 under 35\% motor D and 15 \% motor A dynamic load penetration. 
The first row presents the complete 50~s voltage trajectories for all buses, demonstrating a successful recovery with no evidence of undamped oscillations. From the 3-second post-fault voltage and reactive power measurements, we estimated $K_1$ and $K_2$, estimated the voltage tripping characteristic, and applied EMD as explained in the previous section. 


The second row provides a stability-focused analysis based on local Lyapunov exponents. 
The left subplot shows the FTLEs derived from the IMF components at the system level, whereas the center subplot presents the FSLEs of the residual components for each generator. 
The right subplot includes the artificially generated $S_1$ and $S_2$ tripping signals based on the estimated Voltage caps and the predicted tube. 

The bottom row shows the probability distributions of the divergence factors ($e^{\lambda}$). 
The left subplot presents the IMF-level distribution compared with a shifted–reversed Gompertz reference with $\gamma_2 = 10$, a shift point at~1, and 20~bins distributed over the interval~$[0.0, 1.5]$. Because $0.92 < 2.09$ (the IMF critical value), the index indicates a stable oscillation.
A similar distribution is shown in the center subplot for generator~G7, which exhibits the maximum stability index. 
The right subplot demonstrates how the parameters $\gamma_1$ and the shift point are tuned using the $S_1$ and $S_2$ reference signals.

As seen in the figure, the optimal shift for the Gompertz function corresponding to the residual component $x^*$ is $1.08$, and the optimal $\gamma$ value is $50.8$. The difference between the index values of $S_1$ and $S_2$ is $0.049$, and the critical value is $1.02$. This means that any index value below $1.02 - 0.049/2 \approx 1$ indicates stability, while any value above $1.02 + 0.025 \approx 1.05$ indicates instability. Therefore, critical signals lie within the range of $1.0$ to $1.05$.  Comparing this with the actual index value of the current signal (shown in the middle subplot of the last row), which is $0.3073$, it can be concluded that this generator will not trip. This observation is further validated by the full-time-domain signal displayed in the first row. Note that these threshold values vary from case to case and are calculated online based on the $K_1$ and $K_2$ values, unlike the IMF index values.

Among all generators, this one exhibits the highest index value. The next closest is generator 6. As the dynamic load percentage increases, it was observed that generator 7 trips at a certain loading level while generator 6 remains stable; however, beyond a specific threshold, both generators trip. This confirms the one-to-one characteristic of the index. On the other hand, the index value increases from nearly $0$ to approximately $0.3$ when the dynamic load rises from $0\%$ to $50\%$, and it jumps to more than $0.5$ when the load reaches $60\%$. This indicates that the relationship between the index value and the dynamic load percentage is nonlinear.

Similar to the previous case, Figure~\ref{fig:recovery_case} shows the results for a stressed case with 
55\% of motor D and 30\% of motor C dynamic load penetration under the same fault. 
In the full trajectories, instability emerges around 20--30 seconds after fault clearing, 
driven by slow recovery and eventual collapse. 
The KL divergence values (KL$_{\text{IMF}}=0.93$, KL$_{\text{Residual}}=0.58$) 
show that while oscillatory stability is preserved, the residual index exceeds 
its threshold (KL$_{\text{Residual}}^{critical}=0.325+0.045 \approx 0.375$), indicating recovery-induced 
instability. Importantly, the method detects the instability within 3 seconds, 
well before the collapse occurs.

\begin{figure}[t]
    \centering
    \includegraphics[width=\linewidth]{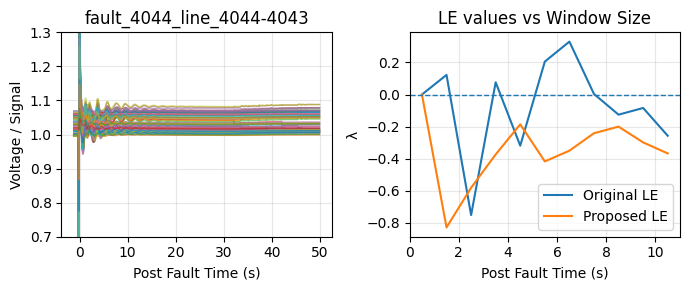}
    \caption{Comparison between the traditional Lyapunov Exponent (LE) and the proposed stability index for a post-fault scenario with 40\% Motor~A composition and +20~dB noise.}
    \label{fig:LE_comparison}
\end{figure}

Fig.~\ref{fig:LE_comparison} compares the performance of the traditional Lyapunov Exponent (LE) with the proposed stability index under a post-fault condition with 40\% Motor~A composition and +20~dB measurement noise. As shown, the traditional LE exhibits significant fluctuation between positive and negative values, leading to inconsistent and delayed stability classification. This behavior aligns with the limitation reported in \cite{koopman_LE2}, where the LE's strong sensitivity to measurement noise prevents it from providing a reliable early indication of system stability. In contrast, the proposed index demonstrates a rapid convergence, consistently identifying the stable response within approximately 0.6~s after fault clearing. This highlights the enhanced robustness and noise tolerance of the proposed method, enabling fast and accurate stability detection in noisy distribution network environments.

\begin{figure}[t]
    \centering
    \includegraphics[width=\linewidth]{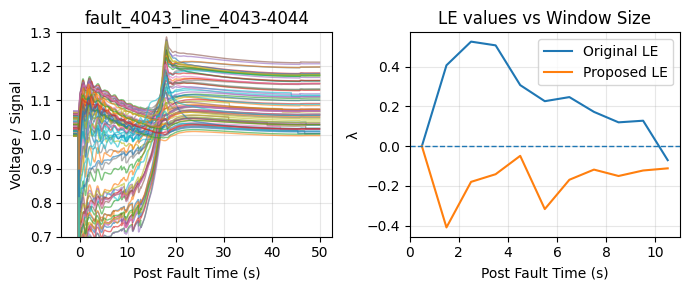}
    \caption{Stability assessment for a post-fault scenario with 15\% Motor~A and 30\% Motor~D, showing slow voltage recovery and oscillatory behavior.}
    \label{fig:LE_comparison_case2}
\end{figure}

Fig.~\ref{fig:LE_comparison_case2} illustrates a stable post-fault case where the voltage trajectory exhibits two interacting behaviors: (i) slow recovery dynamics dominated by Motor~D, and (ii) oscillatory components associated with Motor~A. Even without measurement noise, the traditional Lyapunov Exponent (LE) fails to provide an early stability classification, remaining positive for an extended period before eventually converging. This again demonstrates the LE's limitation in cases where slow dynamic components coexist with oscillatory behavior, as it does not distinguish the contribution of each subsystem to overall stability. In contrast, the proposed index evaluates the oscillatory and slow-recovery components separately, correctly identifying that the oscillatory subsystem is stable while treating the slow voltage recovery through the OEL-related mechanism discussed earlier. As a result, the proposed index classifies the system as stable much earlier than the traditional LE (within 0.6 s). This challenge is not addressed in existing real-time voltage stability assessment methods in the literature, none of which explicitly incorporate slow-recovery dynamics during transient voltage behavior.

\begin{figure*}
    \centering
    \includegraphics[width=\linewidth]{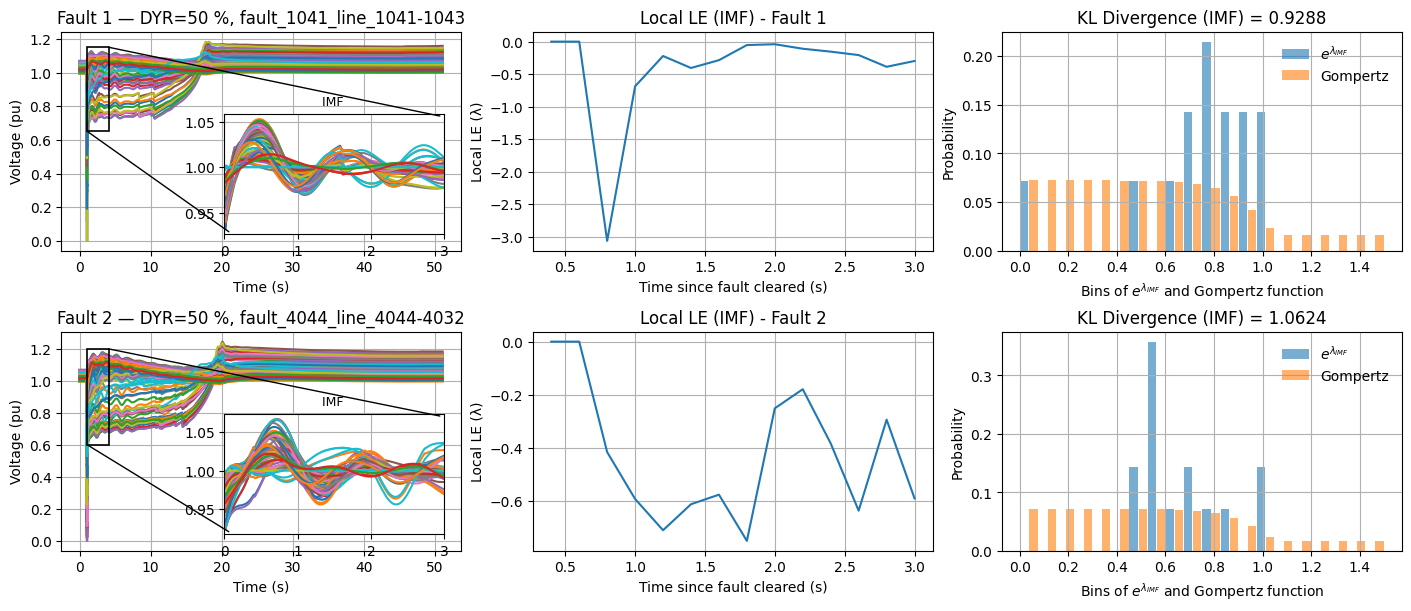}
    \caption{Quantification of the proposed stability index for faults applied at the 132~kV bus (1041) and the 400~kV bus (4044) under 50\% dynamic load penetration. Left column: full 50~s voltage trajectories with the 3~s IMFs highlighted; middle column: local LE evolution; right column: KL divergence between the divergence factor and the reference reversed-shifted Gompertz distribution.}
    \label{fig:quantification_index}
\end{figure*}

Fig.~\ref{fig:quantification_index} presents the quantification of the proposed stability index under two different fault locations 132~kV bus (1041) and  400~kV bus (4044). The computed index values are $0.928$ and $1.062$, respectively, indicating that both cases remain stable; however, the higher value in the 400~kV case reflects a response that is closer to instability. This difference captures the stronger system-level impact of disturbances at higher-voltage buses. As the dynamic load percentage is further increased, instability emerges first for the 400~kV fault case at approximately 63\% penetration, while the 132~kV fault case remains stable and both case instability when dynamic load around 65\%. This confirms that faults at the transmission-level (400~kV) produce more severe voltage divergence tendencies compared to those at the sub-transmission level (132~kV). Together, these results demonstrate that the index captures not only whether the system is stable, but also \emph{how close} the post-fault response is to the boundary of instability, enabling a quantitative and location-sensitive stability assessment.

\section{Conclusion}
This paper presented a data-driven stability index for Short-Term Voltage Stability (STVS) assessment that separates and quantifies the contributions of oscillatory and recovery-related dynamics using Empirical Mode Decomposition (EMD), local Lyapunov exponents, and KL divergence. The proposed formulation overcomes the well-known limitations of traditional Lyapunov-based indicators, enabling fast, noise-robust, and mechanism-aware stability classification. Simulation studies on the Nordic test system demonstrated that the index reliably distinguishes stable and unstable responses, provides early detection within fractions of a second, and produces a quantitative measure of proximity to instability that reflects system strength, fault location, and dynamic load composition.

Future work will focus on extending the index to operate on high-resolution electromagnetic transient (EMT) data, anticipating the evolution of next-generation synchro-wave measurement technologies. This enhancement will allow the framework to assess STVS in the presence of converter-driven dynamics, distinguish between converter-driven instability and voltage collapse mechanisms in EMT-scale signals.

\bibliographystyle{IEEEtran}  
\bibliography{ref}    
\end{document}